# Degradation Mechanism of Perovskite under High Charge Carrier Density Condition

*Guohui Li,*[1] *Huihui Pi,*[1] *Yanfu Wei,*[1] *Bolin Zhou,*[1] *Ya Gao,*[1] *Rong Wen,*[1] *Yuying Hao,*[1] *Han Zhang,*[*,2] *Beng S. Ong*[*,3] *and Yanxia Cui*[*,1]

[1]College of Physics and Optoelectronics, Key Laboratory of Interface Science and Engineering in Advanced Materials, Key Lab of Advanced Transducers and Intelligent Control System of Ministry of Education, Taiyuan University of Technology, Taiyuan 030024, China

[2]Collaborative Innovation Centre for Optoelectronic Science and Technology, Key Laboratory of Optoelectronic Devices and Systems of Ministry of Education and Guangdong Province, College of Physics and Optoelectronic Engineering, Shenzhen Key Laboratory of Micro-Nano Photonic Information Technology, Guangdong Laboratory of Artificial Intelligence and Digital Economy (SZ), Shenzhen University, Shenzhen 518060, China

[3]Department of Chemistry, Research Centre of Excellence for Organic Electronics, Institute of Advanced Materials, Hong Kong Baptist University, Kowloon Tong, Hong Kong, SAR, China

**ABSTRACT**: Extensive studies have focused on degradation of perovskite at low charge carrier density ($<10^{16}$ cm$^{-3}$), but few have surveyed the degradation mechanism at high charge carrier density ($\sim 10^{18}$ cm$^{-3}$). Here, we investigate the degradation mechanisms of perovskite under high charge carrier conditions. Unlike the observations in previous works, we find that MAPbI$_3$ degradation starts at surface defects and progressing from the surface defects towards neighboring regions under high charge carrier density condition. By using PbI$_2$ passivation, the defect-initiated degradation is significantly suppressed and the nanoplatelet degrades in a layer-by-layer way, enabling the MAPbI$_3$ laser sustain for 4500 s ($2.7 \times 10^7$ pulses), which is almost 3 times longer than that of the nanoplatelet laser without passivation. Meanwhile, the PbI$_2$ passivated MAPbI$_3$ nanoplatelet laser with the nanoplatelet cavity displaying a maximum quality factor up to $\sim$7800, the highest reported for all MAPbI$_3$ nanoplatelet cavities. Furthermore, a high stability MAPbI$_3$ nanoplatelet laser that can last for 8500 s ($5.1 \times 10^7$ pulses) is demonstrated based on a dual passivation strategy, by retarding the defect-initiated degradation and surface-initiated degradation, simultaneously. This work provides in-depth insights for understanding the degradation of perovskite at high charge carrier density.

Organic-inorganic hybrid materials, with MAPbI$_3$ as a representative, are of particular interest for developing novel optoelectronic devices working at a wide range of carrier densities such as lasers,[1, 2] light emitting diodes,[3] solar cells,[4, 5] photodetectors,[6] *etc.*, due to their large absorption coefficient, exceptionally low trap-state densities, long charge carrier diffusion lengths, and high charge mobilities.[7] Although preliminary studies on perovskite optoelectronic devices have achieved great success,[1, 8] achieving highly stable perovskite optoelectronic devices is still urgent needed. Degradation of perovskite under low charge carrier density (<$10^{16}$ cm$^{-3}$) condition such as perovskite solar cells has been extensively studied.[9-11] These challenges are more serious in perovskites optoelectronic devices working under high charge carrier density (~$10^{18}$ cm$^{-3}$) condition such as perovskite lasers.

Ascribed to the high charge carrier density, most of the organic and inorganic hybrid perovskite lasers could not sustain more than $10^7$ pulses.[1, 12] For example, the emission intensity of an inkjet-printed MAPbI$_3$ laser on a flexible PET substrate with a nanoimprinted grating in N$_2$ atmosphere dropped to 90% of its initial value after ~$1\times10^6$ pulses.[13] The MAPbI$_3$ laser with a silica microsphere resonator could sustain by $8.6\times10^6$ pulses.[14] Similarly, the solution-processed FAPbBr$_3$ microdisk lasers could work stably for 3000 s ($3\times10^6$ pulses) before dropping to 90% of its initial value.[15] Therefore, investigating the performance degradation mechanism of perovskite lasers under high charge carrier density condition is of critical importance to further improve the operational stability of perovskite lasers.

From one hand, promoting the operating stability of lasers is one of the constant tasks of laser technology.[16] Although room temperature continuous wave perovskite lasers have been reported,[1] one of the major hurdles towards electrically pumped lasers is resistive heating under current injection.[17] On the other hand, improving the stability is of critical importance for achieving electrically pumped perovskite lasers. Until now, great efforts have been made to improve the stability of organic-inorganic perovskites while maintaining the outstanding photophysical properties.[10, 18, 19] The encapsulation strategy has been resorted to improve the perovskite lasing stability. For example, a thin poly-methyl-methacrylate (PMMA) encapsulation layer was applied in a $MAPbI_3$ photonic crystal laser so that the operational stability at a pump intensity of $102.5 \pm 6.4$ $\mu J/cm^2$ being extended from 600 s ($10^5$ pulses) to 6000 s ($10^6$ pulses).[20] By using a CYTOP encapsulation film, a $MAPbI_3$ distributed feedback laser that operated at a pump intensity of 7 $\mu J/cm^2$ could sustain $10^7$ pulses before dropping to 90% of its initial value.[21] It was also demonstrated that the stability of $MAPbI_3$ could be improved by encapsulating with boron nitride flakes.[11] Nevertheless, the degradation mechanism for the hybrid perovskite under high charge carrier density condition remains unknown, and the stability performance of hybrid perovskite lasers is still dissatisfactory.

In this work, by real-time monitoring the emission properties of a $MAPbI_3$ nanoplatelet laser, we find that the gradual degradation of tetragonal $MAPbI_3$ starts from the surface defects and the laser output intensity drops to 90% after ~1200 s ($7.2 \times 10^6$ pulses). Those surface defects on the $MAPbI_3$ nanoplatelets can be effectively passivated by introducing excess $PbI_2$. As a result, the evolution from tetragonal $MAPbI_3$ to $PbI_2$ launches from the crystal surface and the nanoplatelet degrades layer-by-layer, bringing forward the operational stability being extended from 1200 s to 4500 s ($2.7 \times 10^7$ pulses). On the basis of the $PbI_2$ passivated nanoplatelet, we further introduce an additional DBP ($C_{64}H_{36}$) protection film, which can suppress the surface initiated degradation by passivating the surface dangling bonds, thereby dramatically improving the operational stability of the MAPbI$_3$ laser to up to 8500 s ($5.1 \times 10^7$ pulses), which is around 1.89 times as long as that of the MAPbI$_3$ nanoplatelet with only PbI$_2$ passivation. Compared with the initial MAPbI$_3$ nanoplatelets with surface defects, the dual passivation strategy with both PbI$_2$ and DBP enables the MAPbI$_3$ laser sustain 6 times longer, promoting the stability performances of MAPbI$_3$ perovskite lasers significantly. The present passivation strategy of improving the perovskite laser stability paves the way on developing high stability near infrared gain media. In addition, our first attempt on demonstrating the degradation mechanism of the hybrid perovskite crystals under laser pumping might provide in-depth insights for resolving the critical stability hurdle in practical applications of perovskite lasers.

**RESULTS AND DISCUSSION**

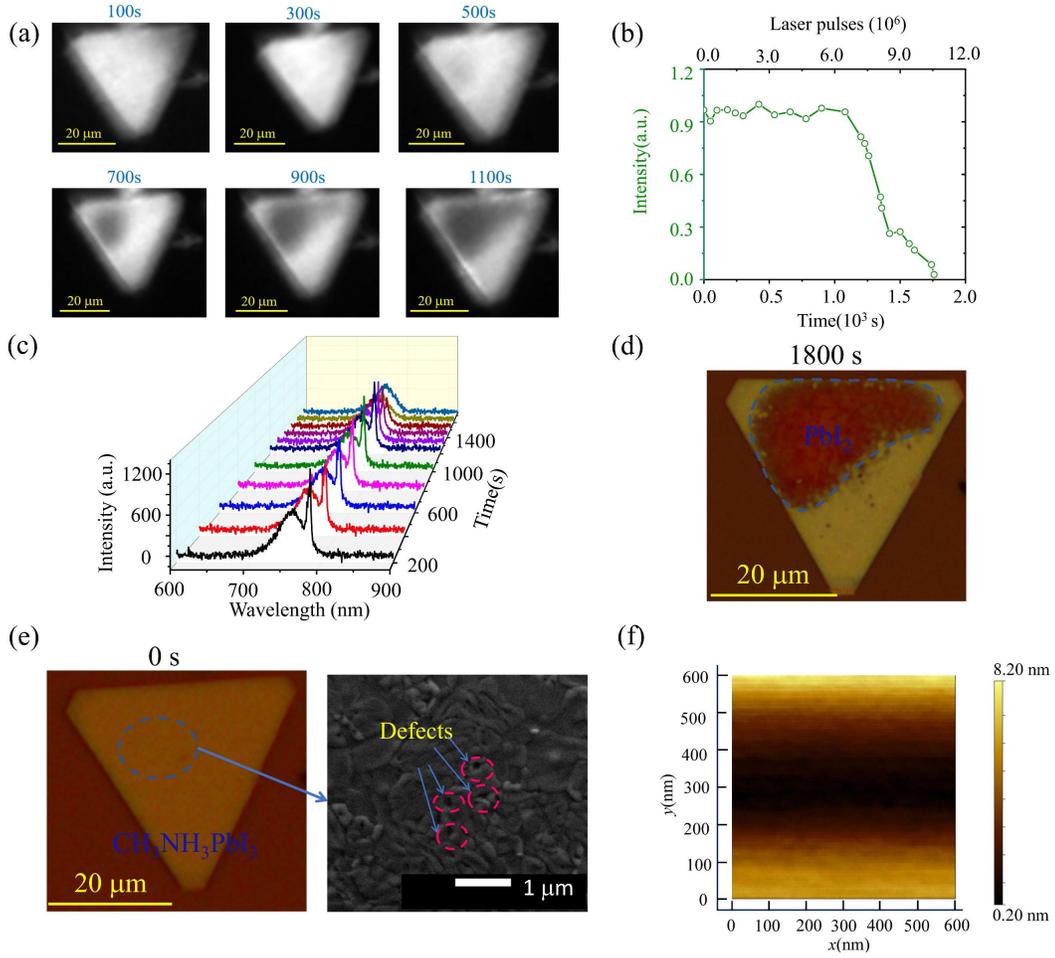

**Figure 1.** (a) Microscopic image of a MAPbI$_3$ laser operating at a pump density of 1.1$P_{th}$ (26.1 μJ/cm$^2$) after working for different times. (b) Lasing stability data of MAPbI$_3$ laser under femtosecond laser pumping with a repetition rate of 6 kHz in ambient air condition. (c) Spectrum evolution of MAPbI$_3$ laser operating at a pump density of 1.1$P_{th}$ (26.1 μJ/cm$^2$) after working for different times. (d) Microscopic image of the nanoplatelet after operating for 1800 s. (e) Microscopic image and SEM image of the initial MAPbI$_3$ nanoplatelet with surface defects. (f) Atomic force microscopic image of the MAPbI$_3$ nanoplatelet shows that its RMS roughness is 2.1 nm.

The MAPbI$_3$ nanoplatelets used in our study were synthesized by the two-step chemical vapor deposition method, that includes the first step of growing PbI$_2$ nanoplatelets and the second step of converting PbI$_2$ nanoplatelets into MAPbI$_3$ nanoplatelets.[2] During the operational stability measurement, we continuously monitored the emission properties and the spectra of the MAPbI$_3$ nanoplatelet laser in ambient air condition with a pumping density of 26.1 µJ/cm$^2$ (1.1$P_{th}$).

As can be seen in Figure 1a, the emission intensity was almost uniform on the surface of the nanoplatelet laser during the first 600 s. After operating for 800 s, the emission intensity on the left side of the nanoplatelet laser started to decrease. Since the measurement takes long time, the emission intensity of the laser operating at a pump density of 1.1$P_{th}$ (26.1 µJ/cm$^2$) during the operating time were measured using an ideaoptics PG2000-Pro spectrometer(See stability characteristic section for more information) as shown in Figure 1b. As can be seen, the laser output intensity as a whole does not change, because more pumping energy can reach lower MAPbI$_3$ layer, that keeps the population inversion $\Delta N$ required for maintaining the output intensity $I \propto \Delta N$ almost unchanged as the upper layer of MAPbI$_3$ degrades; see supporting information for more details. After operating for 1100 s, the emission intensity on a small area on the left side of the nanoplatelet laser decreases dramatically and the area almost becomes dark. After operating for 1200 s (7.2 × 10$^6$ pulses), the dark area increases as shown in Figure 1a and the output intensity of the nanoplatelet laser decreases to 90% of the initial intensity as can be seen in Figure 1b. The operational stability data is in agreement with most of the reported MAPbI$_3$ lasers.[14, 22, 23] After operating for 1400 s, the dark area keeps increasing and the output intensity of the nanoplatelet laser decreases dramatically. The laser dies after working for 1750 s. Besides this nanoplatelet laser, the operational stability of another two unpassivated nanoplatelet lasers has also been measured. The two nanoplatelet lasers can sustain for 1170 s (Figure S1a) and 1200 s (Figure S1b) before output intensity decreases to 90% of their initial value which are consistent with that of the first nanoplatelet laser.

The emission spectrum evolutions of the laser operating at a pump density of $1.1P_{th}$ (26.1 µJ/cm$^2$) during the operating time were also measured by using an ideaoptics PG2000-Pro spectrometer (See stability characteristic section for more information). The pump density corresponds to a charge-carrier density of ~ $8.0 \times 10^{17}$ cm$^{-3}$. From the emission spectrum as shown in Figure 1c, we can see that the intensity of the laser line after operating for 1000 s starts to decrease with the decreasing spontaneous emission intensity. After 1700 s, the spontaneous emission intensity drops down to 50% of its initial intensity and the laser line almost disappears at the same time (see in Figure S2). As can be seen in the microscopic image (Figure 1d) of the MAPbI$_3$ nanoplatelet after operating for 1800 s, some parts of the nanoplatelet have faster degradations and the color of these parts have changed to brown as compared with the yellow color of the rest regions.

From the microscopic image of the initial MAPbI$_3$ nanoplatelet as shown in Figure 1e, it is seen that the nanoplatelet with a thickness of ~ 130 nm (see in Figure S3) initially has a uniform surface and the color of the whole surface is almost the same. However, from the scanning electron microscopy (SEM) images of the MAPbI$_3$ nanoplatelets, some surface defects are found on the surface as can be seen Figure 1e. The atomic force microscopy (AFM) images (Figure 1f) of the nanoplatelets shows that the RMS roughness of the surface is ~2.1 nm. Therefore, the MAPbI$_3$ nanoplatelet under operating condition starts to degrade from the surface defects and progress gradually to neighboring areas as shown in Figure 1a. The corresponding X-ray diffraction (XRD) pattern shows that initially the perovskites nanoplatelets has a pure tetragonal $MAPbI_3$ crystal structure without impurities such as $PbI_2$ (see in Figure S4).   The existence of the small (202), (112), (210), and (221) peaks indicate that the $MAPbI_3$ nanoplatelets are in the room-temperature tetragonal phase.[24] After operating for 1800 s, more than a half of the surface has changed from yellow to brown as can be seen in Figure 1d.   The corresponding XRD pattern shows that (001), (003), and (004) peaks of $PbI_2$ appears after the nanoplatelets operating for 1800 s, confirming that some part of the tetragonal phase $MAPbI_3$ nanoplatelet degrades to $PbI_2$.[24]

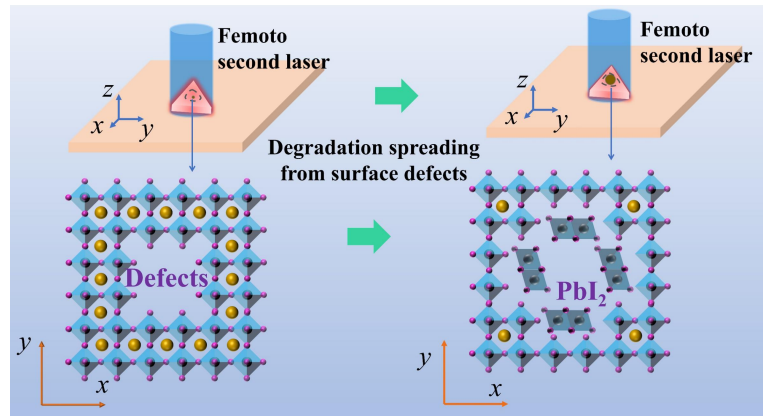

**Figure 2.** Schematic diagram of degradation of perovskite with surface defects that spreads to surrounding regions

The observed phenomenon of $MAPbI_3$ degradation launching from the surface defects deviates from the layer-by-layer degradation theory, which expresses that the thermal-induced-degradation starts from the surface of $MAPbI_3$ as a result of dangling bonds, structure relaxation and charge redistribution on the surface and happens in a sequential layer-by-layer style.[11] A calculation of the transient thermal response of a $MAPbI_3$ nanoplatelet shows that, with a moderate laser pump density of ~17 $\mu J/cm^2$, the transient temperature at the nanoplatelet (see in Figure S6) far exceeds the thermal degradation threshold temperature.[11] It is unquestionable that the MAPbI$_3$ nanoplatelet suffers detrimental thermal-induced-degradation in the experiment. In reality, with respect to the smooth flat surface, the surface defect regions on the surface of the nanoplatelet can form extra dangling bonds on their walls, which initiate new degradation pathways. Since the longer Pb-I-Pb bonds along the [001] direction of MAPbI$_3$ are less resistance to bond breakage than those in the (001) plane,[25] these bonds tend to break first under external stimulus and form dangling bonds. The region with more defects on the nanoplatelet, the faster the speed of the thermal-induced-degradation as shown in Figure 2. Under laser operating conditions, the expansion of the defect region would accelerate the degradation, so a snowball effect is produced. Therefore, ascribed to the existence of surface defects, the degradation proceeds from the inner part to the edge as shown in Figure 2 rather than following the layer-by-layer degradation theory. It is plausible to suppose that reducing the defects can suppress the degradation and making the nanoplatelet lasers operating for longer times.–

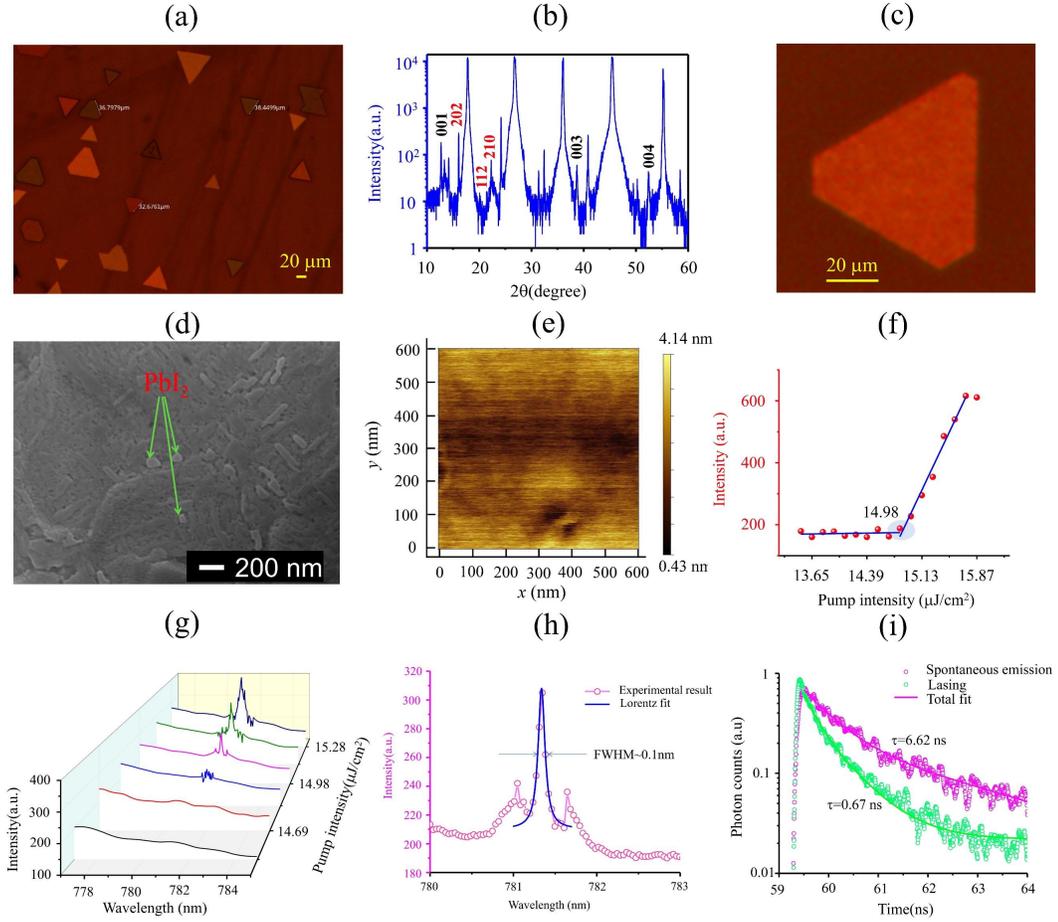

**Figure 3.** (a) Microscopic image of MAPbI$_3$ nanoplatelets on a mica substrate. (b) XRD pattern of the MAPbI$_3$ nanoplatelets. (c) Microscopic image of a MAPbI$_3$ nanoplatelet that used for demonstrating the laser before exposing to a pump laser. (d) SEM image of MAPbI$_3$ nanoplatelet. (e) AFM image of MAPbI$_3$ nanoplatelet shows its RMS roughness is ~ 0.7 nm. (f) Laser output intensity as a function of the pump density. (g) Evolution of emission spectra obtained at different pump densities. (h) Lorentz fitting of a lasing oscillation mode at ≈781.3 nm, gives a FWHM of 0.10 nm, corresponding to a Q factor of 7813. (i) Time-resolved photoluminescence (TRPL) spectra of perovskite nanoplatelet operating at spontaneous emission ($P = 11.12$ μJ/cm$^2$) and laser emission condition ($P = 24.8$ μJ/cm$^2$).

In contrast to fully converting PbI$_2$ to MAPbI$_3$ during the second step of chemical vapor deposition, a certain amount of PbI$_2$ was intentionally reserved to passivate the defects in fabrication of new perovskite nanoplatelets. As shown in Figure 3a, MAPbI$_3$ nanoplatelets with well-defined triangular and hexagonal shape and 100-200 nm thickness and tens of micrometers edge lengths were synthesized.[2] As can be seen in the XRD pattern (Figure 3b), there also exist (001), (003) and (004) peaks of the PbI$_2$ structure in addition to the tetragonal phase MAPbI$_3$ peaks, confirming the excess PbI$_2$ being reserved in the perovskites nanoplatelets. Figure 3c shows the microscopic image of the MAPbI$_3$ nanoplatelet for carrying out the following lasing operation. The perovskite nanoplatelet has a thickness of ~ 180 nm (see in Figure S7). The SEM image in Figure 3d reflects that the surface defects were successfully passivated to a large extent. As can be seen, a newly formed species appeared on the nanoplatelet surface and the new species displayed brighter color as compared with neighboring species as a result of poorer conductivity.[26] According to the XRD pattern as shown in Figure 3b, the species should be PbI$_2$, while the darker films are considered to be perovskite. Here, the unreacted PbI$_2$, without destroying the perovskite crystal structure, is of great helpful for reducing the surface defects in the MAPbI$_3$ nanoplatelets. The AFM image in Figure 3e indicates a RMS roughness of ~ 0.7 nm, confirming that the nanoplatelets have much smoother surfaces supporting the whispering-gallery-mode cavity after passivation.

The influence of excess PbI$_2$ on the laser performance are investigated in the following. The light-in-light-out curve in Figure 3f shows that the emission intensity grows slowly with the increasing pump density below the pump density of ~14.98 μJ/cm$^2$, and then the emission intensity grows very quickly. At a pump intensity of 15.87 μJ/cm$^2$, the emission intensity saturate due to blue shift of center wavelength of the laser.[22] Lasing death did not happen in the measurement. Here, the lasing threshold of ~14.98 µJ/cm² is lower than that of the MAPbI$_3$ nanoplatelet laser without passivation. Since the spectrum has a narrow linewidth which cannot resolved by ideaoptics PG2000-Pro spectrometer, the emission spectrum evolutions of the laser operating at a different pump density were measured by using a Horiba iHR 550 spectrometer (See optical spectrum characterization section for more information). The spectra of the emission light in Figure 3g show that there exists only spontaneous emission below 14.98 µJ/cm². Above the threshold, a narrow laser peak appears and the laser peak increases rapidly with the increasing of pump density. As shown in Figure 3h, separation between adjacent modes is ~0.3 nm which is in agreement with the theoretical value (~0.3 nm) calculated with edge length of the cavity.[2] Lorentz fit of the laser peak at the pump density of 14.98 µJ/cm² shows that the full-width at half-maximum (FWHM) is ~0.1 nm which corresponds to a cavity quality factor Q of 7810, much superior to the values of all reported MAPbI$_3$ nano-laser (the highest reported Q of 3600 belonged to the state-of-the-art MAPbI$_3$ nanowire laser cavity).[22]

We also measured the time-resolved photo-luminescence as shown in Figure 3i. Since MAPbI$_3$ crystals show both fast dynamics and slow dynamics, biexponential fitting were performed to quantify the carrier dynamics. Here, the slow decay component reveals the lifetime of carriers.[27] At a pump density of 11.12 µJ/cm² (below threshold), the PL decay curve shows a long average lifetime of ~6.62 ns. At a pump density of 24.8 µJ/cm² (above the threshold), the PL decay curve shows a short average lifetime of ~0.67 ns. It can be concluded that the lasing threshold has been reduced and quality factor of nanoplatelet cavities has been improved significantly thanks to the reduced surface defects by PbI$_2$ passivation.

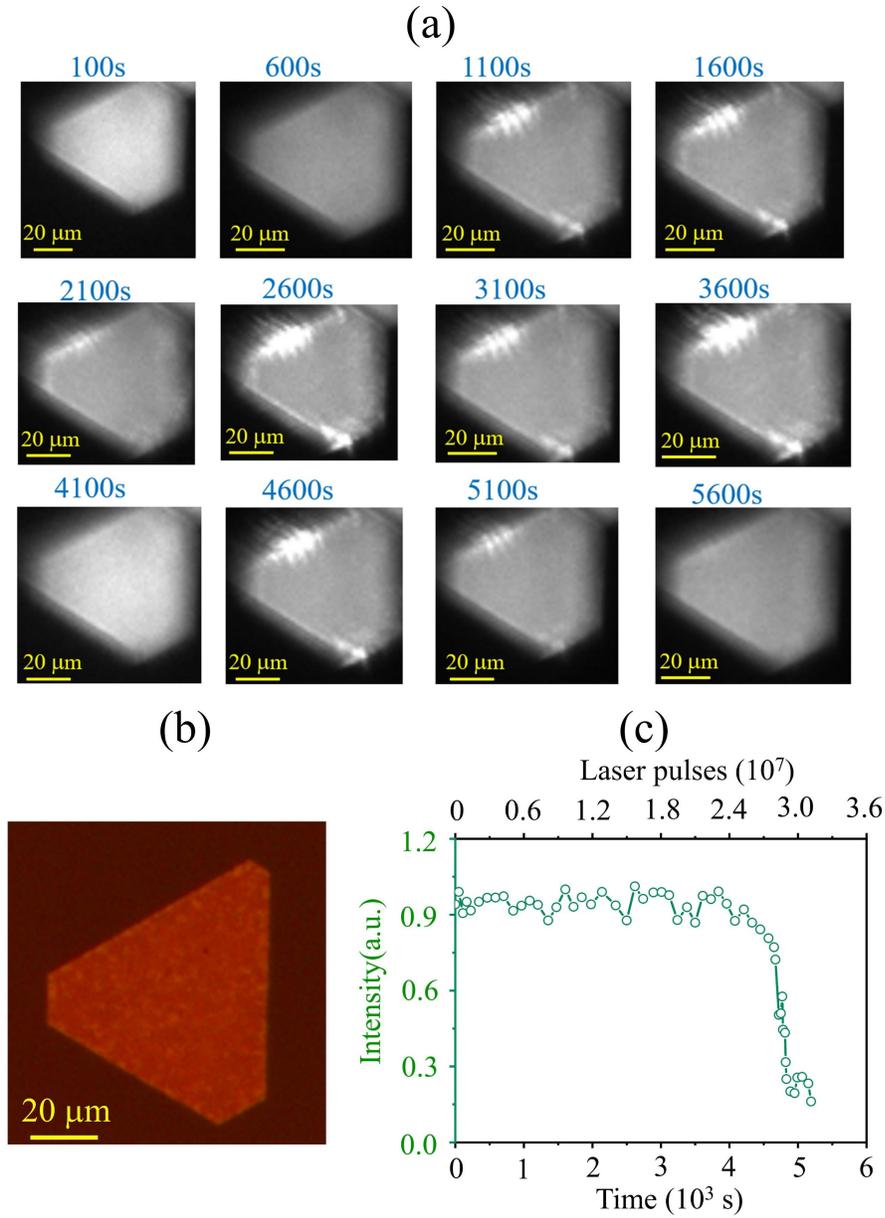

**Figure 4.** (a) Microscopic images of a PbI$_2$ passivated MAPbI$_3$ nanoplatelet laser by operating at a pump density of 1.1$P_{th}$ (16.48 μJ/cm$^2$) for different times. (b) Microscopic image of MAPbI$_3$ nanoplatelet after operating at 1.1$P_{th}$ for 5600 s. (c) Lasing stability

data of PbI$_2$ passivated MAPbI$_3$ nanoplatelet under the femtosecond laser pumping with a repetition rate of 6 kHz in ambient air condition.

Operational stability of the PbI$_2$ passivated MAPbI$_3$ nanoplatelet laser has also been tested under continuous laser pumping with a pumping density of 16.5 μJ/cm$^2$ ($P = 1.1P_{th}$). The pump density corresponds to a charge-carrier density of ~ 5.1×10$^{17}$ cm$^{-3}$. As can be seen in Figure 4a, the laser emission intensity of the PbI$_2$ passivated laser is very stable for 4600 s. After 4600 s, the laser output intensity decreases very rapidly and the emission from surface becomes weak as a whole. After operating for 5600 s, its surface color was still uniform as shown by the microscopic image of the nanoplatelet in Figure 4b. As can be seen in Figure 4c, the monitoring of the laser emission intensity shows that the laser can maintain 90% of the initial intensity after 4500 s (2.7×10$^7$ pulses), which is nearly 3 times longer than that of the MAPbI$_3$ nanoplatelet laser without passivation, and is 2.7 times longer than that of the state of the art MAPbI$_3$ nanowire laser.[22] The emission intensity of the laser operating at a pump density of 1.1$P_{th}$ (16.48 μJ/cm$^2$) during the operating time were also measured by using an ideaoptics PG2000-Pro spectrometer which is capable of long-time stable measurement (See stability characteristic section for more information).

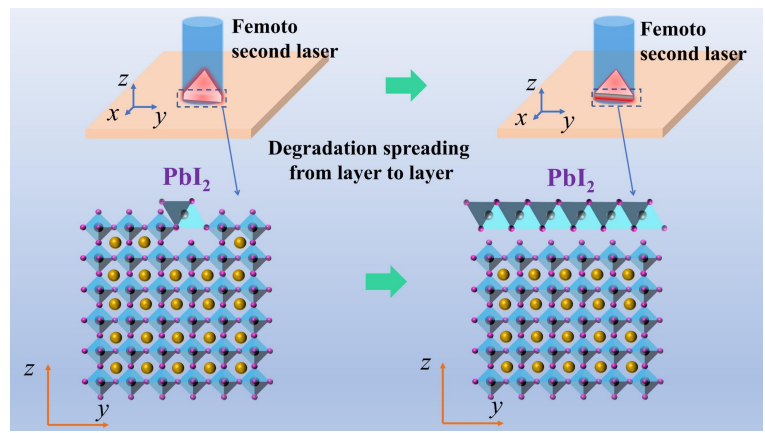

**Figure 5.** Schematic diagram of degradation of perovskite with passivated surface that spreads from layer-to-layer

Thanks to PbI$_2$ passivation, the surface defects are reduced significantly and thereby the surface defects induced degradation are effectively suppressed as shown in Figure 5. Therefore, on the surface of the nanoplatelet, there only exists the dangling bonds triggered thermal decomposition, and correspondingly, the degradation starts from the surface and proceeds layer-by-layer as shown in Figure 5; see supporting information for more details. Besides this PbI$_2$ passivated nanoplatelet laser, the operational stability of another two PbI$_2$ passivated nanoplatelet lasers has also been measured. The two nanoplatelet lasers can sustain for 4400 s (Figure S8a) and 4300 s (Figure S8b) before output intensity decreases to 90% of their initial value which are consistent with that of the first PbI$_2$ passivated nanoplatelet laser.

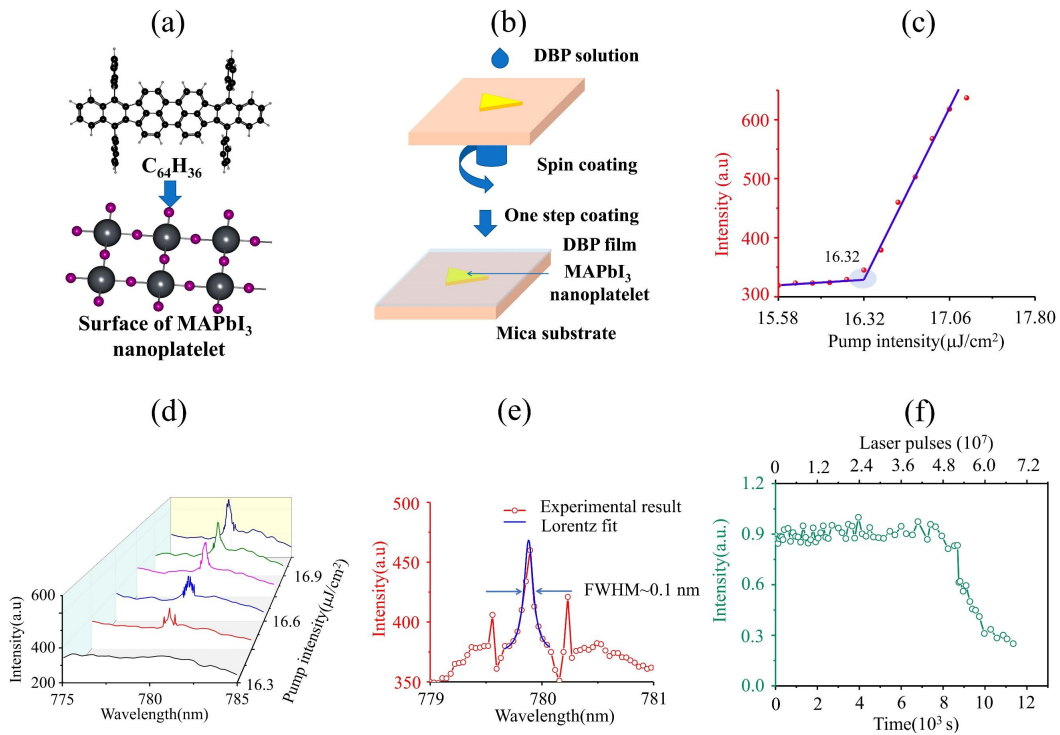

**Figure 6.** (a) Schematic diagram of passivating the surface of MAPbI$_3$ nanoplatelet with DBP (C$_{64}$H$_{36}$). (b) Process illustration of spin-coating a DBP film on the PbI$_2$ passivated MAPbI$_3$ nanoplatelet. (c) Laser output intensity as a function of pump density. (d) Evolution of emission spectra obtained at different pump densities. (e) Lorentz fitting of a lasing oscillation mode at ≈779.9 nm, gives a FWHM of 0.10 nm, corresponding to a Q factor of 7799. (f) Lasing stability data of the dual passivation processed MAPbI$_3$ nanoplatelet laser under the femtosecond laser pumping with a repetition rate of 6 kHz in ambient air condition. Dual passivation refers to PbI$_2$ passivation and DBP passivation.

Next, we optimized the operational stability of PbI$_2$ passivated MAPbI$_3$ nanoplatelet laser by introducing an additional encapsulation layer to passivate the surface of the nanoplatelet. The surface-imitated layer-by-layer degradation of MAPbI$_3$ is considered to be caused by Pb and I dangling bonds on the MAPbI$_3$ surface. Hydrogen and pseudo-hydrogen atoms are supposed to provide an ideal passivation to pair the electron in the dangling bonds on the surface of semiconductor nano-structures.[28, 29] DBP (C$_{64}$H$_{36}$) is a promising material for improving the performances of perovskite optoelectronic devices such as solar cells and light emitting diodes.[30, 31]

To suppress the surface-imitated degradation of perovskite nanoplatelets, we employed a thin DBP film as the encapsulation layer on newly synthesized PbI$_2$ passivated MAPbI$_3$ nanoplatelet surface to form a DBP-MAPbI$_3$-mica heterostructure as shown in Figure 6a. The DBP film was spin-coated on the surface of MAPbI$_3$ nanoplatelets on mica substrate as shown in Figure 6b. After coating the DBP film, the MAPbI$_3$ nanoplatelets on mica substrate (Figure S9a) become darker as compared with the uncoated MAPbI$_3$ nanoplatelets on mica substrate (Figure S9b). Without passivation, the MAPbI$_3$ surface with Pb and I dangling bonds is more susceptible to degrade. As shown in Figure S10, the yellow nanoplatelet degrades severely for 48 h in ambient air condition. Instead, the DBP encapsulated nanoplatelet can keep in ambient air condition for more than 120 h as can be seen in Figure S11. That is because, with DBP encapsulation, the H$^-$ in the $C_{64}H_{36}$ interact with perovskite surface dangling bonds, which effectively reduces the surface activity and enables a highly stable MAPbI$_3$ nanoplatelet.

The lasing performances of the DBP encapsulated MAPbI$_3$ nanoplatelet laser are shown in Figure 6. It is found that the lasing threshold (~16.32 μJ/cm$^2$) of MAPbI$_3$ nanoplatelet laser is slightly increased by DBP encapsulation as shown in Figure 6c, which might be induced by light absorption of DBP. Since the spectrum has a narrow linewidth which cannot resolved by ideaoptics PG2000-Pro spectrometer, the emission spectrum evolutions of the laser operating at a different pump density were also measured by using a Horiba iHR 550 spectrometer (See optical spectrum characterization section for more information). The spectra of the emission light in Figure 6d show that there exists only spontaneous emission below 16.32 μJ/cm$^2$. Above the threshold, a narrow laser peak appears and the laser peak increases rapidly with the increasing of pump density. As can be seen in Figure 6e, Lorentz fit of the laser peak at the pump density of 16.47 μJ/cm$^2$ shows that the FWHM is ~0.1 nm which corresponds to a cavity quality factor Q of ~ 7799.

We then performed the operational stability test of the obtained stable MAPbI$_3$ nanoplatelet at a pump density of 1.1$P_{th}$ (~17.95 μJ/cm$^2$) at room temperature in ambient air condition. The pump density corresponds to a charge-carrier density of ~ $5.5 \times 10^{17}$

cm$^{-3}$. Since the measurement takes long time, the emission intensity of the laser was measured by using an ideaoptics PG2000-Pro spectrometer (See stability characteristic section for more information). As can be seen in Figure 6f, it shows that the dual passivation processed MAPbI$_3$ nanoplatelet laser have considerably improved operational stability. The output intensity of the dual passivation processed laser keeps 90 % of the initial value for longer than 8500 s (5.1×10$^7$ pulses), which is around 1.89 times as long as that of the MAPbI$_3$ nanoplatelet with only PbI$_2$ passivation. Compared with the initial unpassivated MAPbI$_3$ nanoplatelets with surface defects, the dual passivation strategy enables the MAPbI$_3$ laser sustains 6 times longer, outperforming the performances of all reported hybrid perovskite lasers. Its operational stability is even better than some of the all inorganic CsPbBr$_3$ lasers.[32, 33] This result confirms that the rich hydrogen atoms contained in the DBP molecules can provide effective passivation of the electron in the dangling bonds on the surface of MAPbI$_3$ nanoplatelets. Besides this dual passivation processed nanoplatelet laser, the operational stability of another two dual passivation processed nanoplatelet lasers has also been measured. The two nanoplatelet lasers can sustain for 8290 s (Figure S12a) and 8390 s (Figure S12b) before output intensity decreases to 90% of their initial value which are consistent with that of the first dual passivation processed nanoplatelet laser.

The average operation time of unpassivated (sample A), PbI$_2$ passivated (sample B) and dual passivation processed nanoplatelet lasers (sample C) under femtosecond laser pumping with a repetition rate of 6 kHz in ambient air condition are 1190 s, 4400 s and 8450 s (see Figure S13), respectively. It can be seen that the average operation time of PbI$_2$ passivated nanoplatelet lasers is more than three times longer than that of the unpassivated nanoplatelet lasers. Through dual passivation processing, the average operation time of nanoplatelet lasers is improved more than seven times as compared with that of the unpassivated nanoplatelet lasers.

**CONCLUSION**

We investigate the emission intensity and spectrum changes of perovskite nanoplatelet lasers under operational condition. We reveal that degradation of perovskite under high charge carrier density condition stems from the thermal-induced-degradation which starts at the surface defects on the surface of MAPbI$_3$ and then progresses towards the neighboring regions. Excess PbI$_2$ can be employed to passivate the surface defects. After passivation, the nanoplatelet degrades from layer-to-layer. As a result, the PbI$_2$ passivated nanoplatelet laser can sustain for 4500 s ($2.7 \times 10^7$ pulses), which is more than 3 times longer than the nanoplatelet laser without passivation. It has been demonstrated that the PbI$_2$ passivated nanoplate laser has a threshold as low as 14.98 μJ/cm$^2$ and a cavity quality factor up to ~7810. To further retard the surface-initiated degradation, an additional DBP film has been utilized as a protection layer on the PbI$_2$ passivated MAPbI$_3$ nanoplatelet. The DBP encapsulated nanoplatelet show considerably improved operational stability which can last for 8500 s ($5.1 \times 10^7$ pulses) until it falls to 90% of its initial intensity. There findings pave the way for enhancing long-term operational stability of perovskite optoelectronic devices working at high charge carrier density condition.

## METHODS

**Synthesis of Perovskite NPLs.** $PbI_2$ (99.999%, Alfa) was used as a single source and placed into a quartz tube mounted on a single zone furnace (CY scientific instrument, CY-O1200-1L) at a room temperature of 18 °C. The fresh-cleaved muscovite mica substrate was pre-cleaned with acetone and placed in the downstream region inside the quartz tube. The quartz tube was first evacuated to 0.1 Pa, followed by a 30 sccm flow of high purity Ar premixed with 10% $H_2$ gas. The temperature and pressure inside the quartz tube were set and stabilized at 380 °C and 0.12 MPa for $PbI_2$. The synthesis of $PbI_2$ was completed within 14 min, and the furnace was allowed to cool naturally to room temperature. Then, pre-grown lead halide nanoplatelets were thermally intercalated with MAI (Xi'an Polymer light technology) in a fresh quartz tube. The mica substrate with nanoplatelets was placed in the downstream region, while the MAI powder was placed in the center of the tube. The intercalation was carried out at 120 °C at a pressure of 0.11 MPa with a 34-sccm flow of high purity Ar for 200 min to convert the lead halides to perovskites completely. For $PbI_2$ passivation, the intercalation was carried out for 170 min to keep parts of lead iodide for passivation.

**Fabrication of the DBP film.** 0.002 g DBP (99%, Han Feng) was first fully dissolved in 1 mL chlorobenzene (Sigma). After filtration, 20 μL DBP solution was spin-coated on the surface of the perovskite nanoplatelets at 4500 rpm for 30 s in an $N_2$ filled glovebox. The film formed after 2 min.

**Image and phase characterizations.** The optical images of $MAPbI_3$ nanostructures were obtained on a Nikon LV150 optical microscope. The AFM images were collected on an FM-Nanoview 1000 AFM (FSM Precision) which samples 512 points separately in x and y direction. The XRD data were acquired on a DX-2700 diffractometer (Dandong Haoyuan) by using a sampling time of 0.1 s. The SEM images were obtained at an accelerating voltage of 5.0 kV by using a JEOL JSM-IT500 scanning electron microscope.

**Optical spectrum characterization.** We carried out optically pumped lasing measurements on a home-built microscope setup. The 343 nm excitation pulses were generated by frequency tripling the 1028 nm output (with a BBO crystal) from a Light Conversion Carbide Femtosecond laser (290 fs, 6 kHz, 1028 nm). The pumping source was focused onto samples *via* an uncoated convex lens (Focal length: 20 cm, Transmittance: 80%). To ensure uniform energy injection, the laser spot diameter was focused to ~ 107 μm. The transmitted emission was collected through a 20× objective lens (Olympus, Numerical Aperture: 0.4). Half of the emission signals were imaged on a camera (Hamamatsu, C11440-36U). The other half of the emission signal from a single nanoplatelet was collected into an optical fiber with core diameter of 600 μm and analyzed using a Horiba iHR 550 equipped with a symphony CCD head. Each spectrum was obtained through a single measurement. The CCD head has a E2V manufactured 2048 × 512 pixel Back Illuminated Visible CCD chip and was cooled to 140 K with liquid $N_2$. The spectrometer can work stably for 4 h after filled with liquid $N_2$. A 1200 g/mm, 500 nm blazed, 76 mm × 76 mm, and ion-etched holographic diffraction grating and the entrance slit of 50 μm were used in the measurement. The spectral resolution of the spectrometer is ~ 0.04 nm. The emission was time-resolved by using a TCSPC module (Picoquant, PicoHarp 300) and a SPAD detector (MPD, PD-100-CTE) with an instrument response function of 30 ps (FWHM).

**Stability characterization.** The emission intensity from a single nanoplatelet was monitored using an ideaoptics PG2000-Pro spectrometer with a wavelength resolution (FWHM) of 0.3 nm in the range 700-900 nm. Since the spectrometer does not require cooling liquid, it can work stably for longer times. For spectral range of 200-1100 nm, the ideaoptics PG2000 spectrometer with a wavelength resolution (FWHM) of 1.3 nm was used.

## AUTHOR INFORMATION

**Corresponding Author**

*E-mal:yanxiacui@gmail.com(Y. Cui), hzhang@szu.edu.cn(H. Zhang), bong@hkbu.edu.hk (B. S. Ong)

**Author Contributions**

The manuscript was written through contributions of all authors. All authors have given approval to the final version of the manuscript.

**Notes**

The authors declare no competing financial interest.

## ACKNOWLEDGMENTS

This work was supported by the National Natural Science Foundation of China (61922060, 61775156, 61961136001 and 61875138). The commercialization cultivation program of Shanxi college research findings (2020CG013). YC also acknowledges support from Key Research and Development (International Cooperation) Program of Shanxi Province (201803D421044), Henry Fok Education Foundation Young Teachers fund, and Platform and Base Special Project of Shanxi Province (201805D131012-3). HZ also acknowledges support from the Nature Science Foundation of Guangdong Province (2018A030313401), Shenzhen Nanshan District Pilotage Team Program (LHTD20170006), and the Science and Technology Innovation Commission of Shenzhen (JCYJ20170811093453105, JCYJ20170818141519879).

**Supporting Information Available:**

Lasing stability data of another two unpassivated $MAPbI_3$ lasers; Emission spectra of the unpassivated $MAPbI_3$ nanoplatelet lasers; AFM image of the edge of the unpassivated $MAPbI_3$ nanoplatelet; XRD patterns of $MAPbI_3$ nanoplatelets on mica substrate; Transient thermal response of a $MAPbI_3$ nanoplatelet; AFM image of the edge of $PbI_2$ passivated $MAPbI_3$ nanoplatelet; Lasing stability data of another two $PbI_2$ passivated $MAPbI_3$ lasers; Average operation time of unpassivated, $PbI_2$ passivated and dual passivation processed nanoplatelet lasers.

## REFERENCES

1. Qin, C.; Sandanayaka, A. S. D.; Zhao, C.; Matsushima, T.; Zhang, D.; Fujihara, T.; Adachi, C., Stable room-temperature continuous-wave lasing in quasi-2D perovskite films. *Nature* **2020,** *585* (7823), 53-57.
2. Li, G.; Che, T.; Ji, X.; Liu, S.; Hao, Y.; Cui, Y.; Liu, S., Record-Low-Threshold Lasers Based on Atomically Smooth Triangular Nanoplatelet Perovskite. *Adv. Funct. Mater.* **2019,** *29* (2), 1805553.
3. Veldhuis, S. A.; Boix, P. P.; Yantara, N.; Li, M.; Sum, T. C.; Mathews, N.; Mhaisalkar, S. G., Perovskite Materials for Light-Emitting Diodes and Lasers. *Advanced Materials* **2016,** *28* (32), 6804-6834.
4. Green, M.; Dunlop, E.; Hohl-Ebinger, J.; Yoshita, M.; Kopidakis, N.; Hao, X., Solar cell efficiency tables (version 57). *Progress in Photovoltaics: Research and Applications* **2021,** *29* (1), 3-15.
5. Chen, Y.; Zuo, X.; He, Y.; Qian, F.; Zuo, S.; Zhang, Y.; Liang, L.; Chen, Z.; Zhao, K.; Liu, Z.; Gou, J.; Liu, S., Dual Passivation of Perovskite and SnO2 for High-Efficiency MAPbI3 Perovskite Solar Cells. *Advanced Science* **2021,** *8* (5), 2001466.
6. Li, G.; Gao, R.; Han, Y.; Zhai, A.; Liu, Y.; Tian, Y.; Tian, B.; Hao, Y.; Liu, S.; Wu, Y.; Cui, Y., High detectivity photodetectors based on perovskite nanowires with suppressed surface defects. *Photonics Research* **2020,** *8* (12), 1862-1874.
7. Manser, J. S.; Christians, J. A.; Kamat, P. V., Intriguing Optoelectronic Properties of Metal Halide Perovskites. *Chem. Rev.* **2016,** *116* (21), 12956-13008.
8. Jia, Y.; Kerner, R. A.; Grede, A. J.; Rand, B. P.; Giebink, N. C., Continuous-wave lasing in an organic–inorganic lead halide perovskite semiconductor. *Nature Photonics* **2017,** *11* (12), 784-788.
9. Guo, R.; Han, D.; Chen, W.; Dai, L.; Ji, K.; Xiong, Q.; Li, S.; Reb, L. K.; Scheel, M. A.; Pratap, S.; Li, N.; Yin, S.; Xiao, T.; Liang, S.; Oechsle, A. L.; Weindl, C. L.; Schwartzkopf, M.; Ebert, H.; Gao, P.; Wang, K.; Yuan, M.; Greenham, N. C.; Stranks, S. D.; Roth, S. V.; Friend, R. H.; Müller-Buschbaum, P., Degradation mechanisms of perovskite solar cells under vacuum and one atmosphere of nitrogen. *Nature Energy* **2021,** *6* (10), 977-986.
10. Leijtens, T.; Eperon, G. E.; Noel, N. K.; Habisreutinger, S. N.; Petrozza, A.; Snaith, H. J., Stability of Metal Halide Perovskite Solar Cells. *Advanced Energy Materials* **2015,** *5* (20), 1500963.
11. Fan, Z.; Xiao, H.; Wang, Y.; Zhao, Z.; Lin, Z.; Cheng, H.-C.; Lee, S.-J.; Wang, G.; Feng, Z.; Goddard, W. A.; Huang, Y.; Duan, X., Layer-by-Layer Degradation of Methylammonium Lead Tri-iodide Perovskite Microplates. *Joule* **2017,** *1* (3), 548-562.
12. Sun, W.; Liu, Y.; Qu, G.; Fan, Y.; Dai, W.; Wang, Y.; Song, Q.; Han, J.; Xiao, S., Lead halide perovskite vortex microlasers. *Nat Commun* **2020,** *11* (1), 4862.
13. Mathies, F.; Brenner, P.; Hernandez-Sosa, G.; Howard, I. A.; Paetzold, U. W.; Lemmer, U., Inkjet-printed perovskite distributed feedback lasers. *Opt. Express* **2018,** *26* (2), A144-A152.

14. Sutherland, B. R.; Hoogland, S.; Adachi, M. M.; Wong, C. T. O.; Sargent, E. H., Conformal Organohalide Perovskites Enable Lasing on Spherical Resonators. *ACS Nano* **2014,** *8* (10), 10947-10952.
15. Li, X.; Wang, K.; Chen, M.; Wang, S.; Fan, Y.; Liang, T.; Song, Q.; Xing, G.; Tang, Z., Stable Whispering Gallery Mode Lasing from Solution-Processed Formamidinium Lead Bromide Perovskite Microdisks. *Advanced Optical Materials* **2020,** *8* (15), 2000030.
16. Zhang, Q.; Shang, Q.; Su, R.; Do, T. T. H.; Xiong, Q., Halide Perovskite Semiconductor Lasers: Materials, Cavity Design, and Low Threshold. *Nano Lett.* **2021,** *21* (5), 1903-1914.
17. Sutherland, B. R.; Sargent, E. H., Perovskite photonic sources. *Nature Photonics* **2016,** *10*, 295.
18. Li, G.; Chen, K.; Cui, Y.; Zhang, Y.; Tian, Y.; Tian, B.; Hao, Y.; Wu, Y.; Zhang, H., Stability of Perovskite Light Sources: Status and Challenges. *Advanced Optical Materials* **2020,** *8* (6), 1902012.
19. Fu, Q.; Tang, X.; Huang, B.; Hu, T.; Tan, L.; Chen, L.; Chen, Y., Recent Progress on the Long-Term Stability of Perovskite Solar Cells. *Advanced Science* **2018,** *5* (5), 1700387.
20. Chen, S.; Roh, K.; Lee, J.; Chong, W. K.; Lu, Y.; Mathews, N.; Sum, T. C.; Nurmikko, A., A Photonic Crystal Laser from Solution Based Organo-Lead Iodide Perovskite Thin Films. *ACS Nano* **2016,** *10* (4), 3959-3967.
21. Whitworth, G. L.; Harwell, J. R.; Miller, D. N.; Hedley, G. J.; Zhang, W.; Snaith, H. J.; Turnbull, G. A.; Samuel, I. D. W., Nanoimprinted distributed feedback lasers of solution processed hybrid perovskites. *Opt. Express* **2016,** *24* (21), 23677-23684.
22. Zhu, H.; Fu, Y.; Meng, F.; Wu, X.; Gong, Z.; Ding, Q.; Gustafsson, M. V.; Trinh, M. T.; Jin, S.; Zhu, X. Y., Lead halide perovskite nanowire lasers with low lasing thresholds and high quality factors. *Nature Materials* **2015,** *14*, 636.
23. Whitworth, G. L.; Harwell, J. R.; Miller, D. N.; Hedley, G. J.; Zhang, W.; Snaith, H. J.; Turnbull, G. A.; Samuel, I. D., Nanoimprinted distributed feedback lasers of solution processed hybrid perovskites. *Opt. Express* **2016,** *24* (21), 23677-23684.
24. Ha, S. T.; Liu, X.; Zhang, Q.; Giovanni, D.; Sum, T. C.; Xiong, Q., Synthesis of Organic–Inorganic Lead Halide Perovskite Nanoplatelets: Towards High-Performance Perovskite Solar Cells and Optoelectronic Devices. *Advanced Optical Materials* **2014,** *2* (9), 838-844.
25. Brivio, F.; Frost, J. M.; Skelton, J. M.; Jackson, A. J.; Weber, O. J.; Weller, M. T.; Goñi, A. R.; Leguy, A. M. A.; Barnes, P. R. F.; Walsh, A., Lattice dynamics and vibrational spectra of the orthorhombic, tetragonal, and cubic phases of methylammonium lead iodide. *Physical Review B* **2015,** *92* (14), 144308.
26. Chen, Q.; Zhou, H.; Song, T.-B.; Luo, S.; Hong, Z.; Duan, H.-S.; Dou, L.; Liu, Y.; Yang, Y., Controllable Self-Induced Passivation of Hybrid Lead Iodide Perovskites toward High Performance Solar Cells. *Nano Lett.* **2014,** *14* (7), 4158-4163.
27. Shi, D.; Adinolfi, V.; Comin, R.; Yuan, M.; Alarousu, E.; Buin, A.; Chen, Y.; Hoogland, S.; Rothenberger, A.; Katsiev, K.; Losovyj, Y.; Zhang, X.; Dowben, P. A.; Mohammed, O. F.; Sargent, E. H.; Bakr, O. M., Low trap-state density and long carrier diffusion in organolead trihalide perovskite single crystals. *Science* **2015,** *347* (6221), 519.
28. Li, J.; Wei, S.-H.; Wang, L.-W., Stability of the DX$^{-1}$ Center in GaAs Quantum Dots. *Phys. Rev. Lett.* **2005,** *94* (18), 185501.
29. Li, J.; Wang, Comparison between Quantum Confinement Effects of Quantum Wires and Dots. *Chem. Mater.* **2004,** *16* (21), 4012-4015.
30. Ding, S.; Li, S.; Sun, Q.; Wu, Y.; Liu, Y.; Li, Z.; Cui, Y.; Wang, H.; Hao, Y.; Wu, Y., Enhanced performance of perovskite solar cells by the incorporation of the luminescent small molecule DBP: perovskite absorption spectrum modification and interface engineering. *Journal of Materials Chemistry C* **2019,** *7* (19), 5686-5694.
31. Kirchhuebel, T.; Gruenewald, M.; Sojka, F.; Kera, S.; Bussolotti, F.; Ueba, T.; Ueno, N.; Rouillé, G.; Forker, R.; Fritz, T., Self-Assembly of Tetraphenyldibenzoperiflanthene (DBP) Films on Ag(111) in the Monolayer Regime. *Langmuir* **2016,** *32* (8), 1981-1987.
32. Tang, B.; Dong, H.; Sun, L.; Zheng, W.; Wang, Q.; Sun, F.; Jiang, X.; Pan, A.; Zhang, L., Single-Mode Lasers Based on Cesium Lead Halide Perovskite Submicron Spheres. *ACS Nano* **2017,** *11* (11), 10681-10688.
33. Wang, Y.; Li, X.; Song, J.; Xiao, L.; Zeng, H.; Sun, H., All-Inorganic Colloidal Perovskite Quantum Dots: A New Class of Lasing Materials with Favorable Characteristics. *Adv. Mater.* **2015,** *27* (44), 7101-7108.

# Supporting information

**Degradation Mechanism of Perovskite under High Charge Carrier Density Condition**

*Guohui Li,*[1] *Huihui Pi,*[1] *Yanfu Wei,*[1] *Bolin Zhou,*[1] *Ya Gao,*[1] *Rong Wen,*[1] *Yuying Hao,*[1] *Han Zhang,*[*,2] *Beng S. Ong*[*,3] *and Yanxia Cui*[*,1]

[1]College of Physics and Optoelectronics, Key Laboratory of Interface Science and Engineering in Advanced Materials, Key Lab of Advanced Transducers and Intelligent Control System of Ministry of Education, Taiyuan University of Technology, Taiyuan 030024, China
[2]Collaborative Innovation Centre for Optoelectronic Science and Technology, Key Laboratory of Optoelectronic Devices and Systems of Ministry of Education and Guangdong Province, College of Physics and Optoelectronic Engineering, Shenzhen Key Laboratory of Micro-Nano Photonic Information Technology, Guangdong Laboratory of Artificial Intelligence and Digital Economy (SZ), Shenzhen University, Shenzhen 518060, China

[3]Department of Chemistry, Research Centre of Excellence for Organic Electronics, Institute of Advanced Materials, Hong Kong Baptist University, Kowloon Tong, Hong Kong, SAR, China

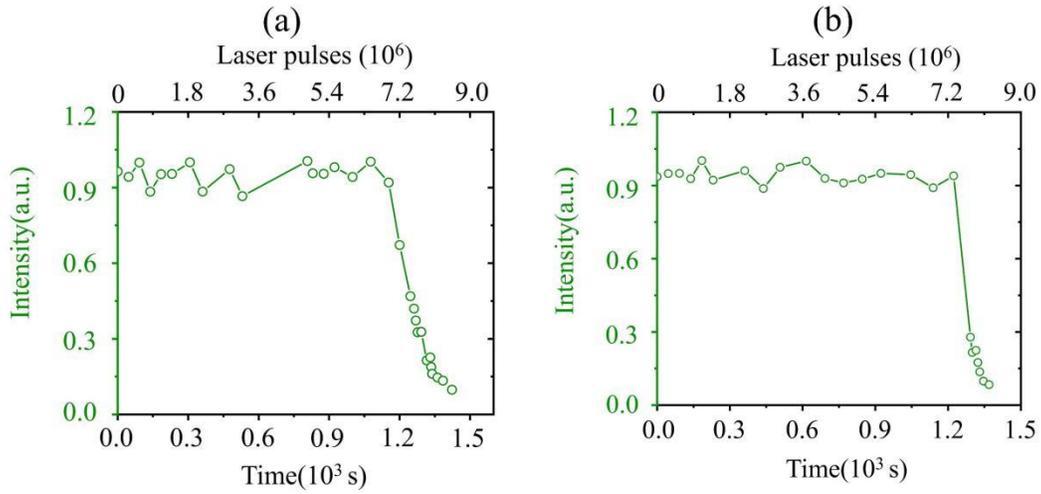

**Figure S1.** Lasing stability data of another two unpassivated MAPbI$_3$ lasers (a) and (b) under femtosecond laser pumping with a repetition rate of 6 kHz in ambient air condition.

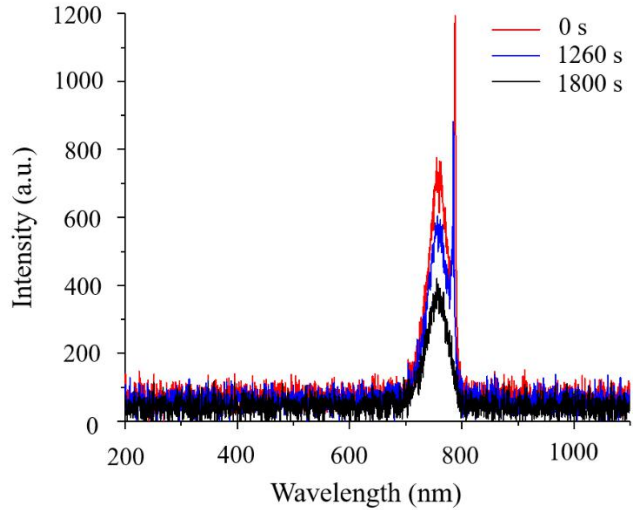

**Figure S2.** Emission spectra of the unpassivated MAPbI$_3$ nanoplatelet lasers after operating for different times measured using ideaoptics PG2000 spectrometer (See stability characteristic section for more information).

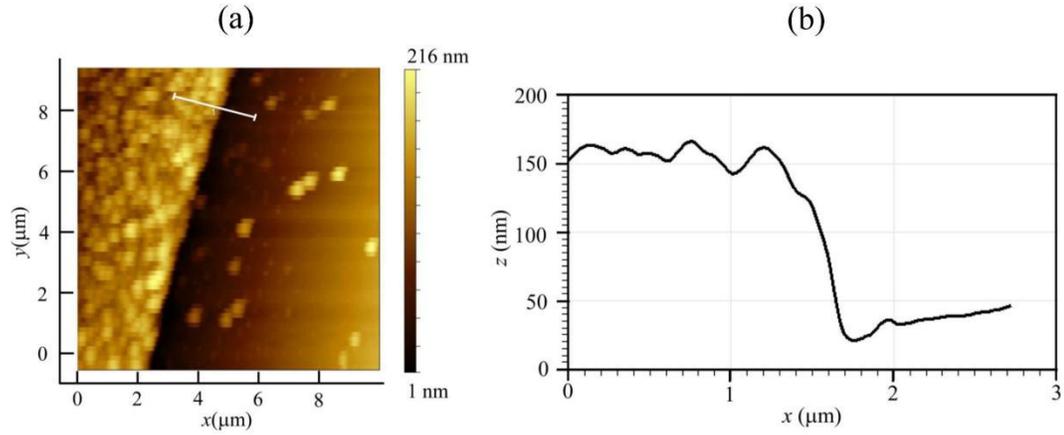

**Figure S3.** AFM image of the edge of the unpassivated MAPbI$_3$ nanoplatelet (a) and the corresponding cross view for showing the thickness (b).

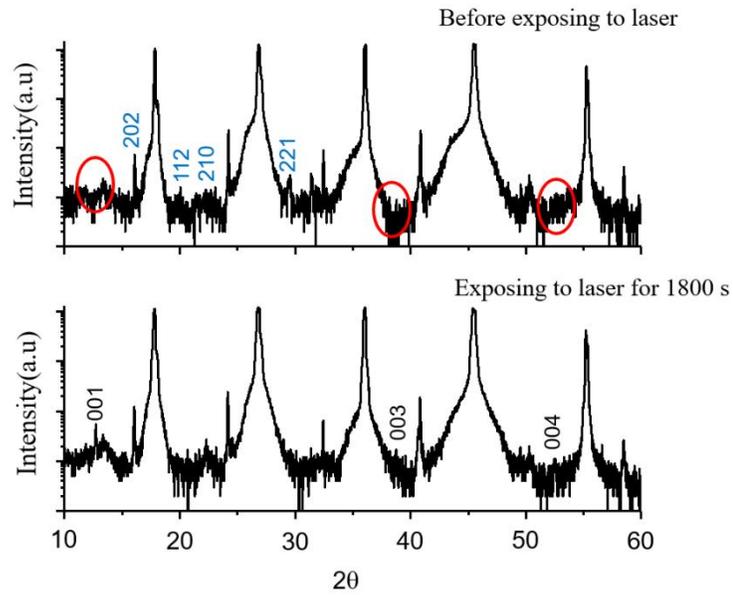

**Figure S4.** XRD patterns of MAPbI$_3$ nanoplatelets on mica substrate before and after exposing to the pump laser for 1800 s.

1. Relations for population inversion related laser output

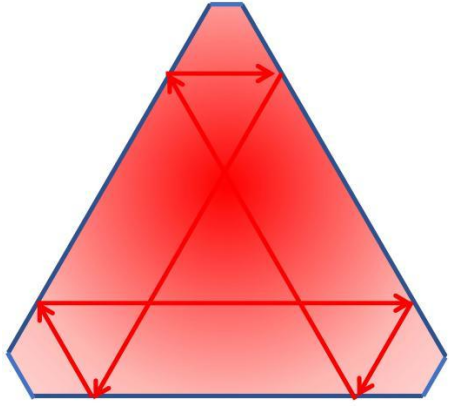

**Figure S5.** Schematic diagram of the light path in a MAPbI$_3$ nanoplatelet.

A light wave traveling through a gain medium can be expressed as

$$E(z,t) = E_0 \exp[i(\omega t - kz)] \cdot \exp[i\Delta k + \gamma(\omega)/2)z] \qquad (S1)$$

where $\Delta kz$ is the phase change of the light wave, γ is the laser gain

$$\gamma(\omega) = -\frac{\lambda^2 \Delta N}{8\pi n^2 \tau_{sp}} g(v) \qquad (S2)$$

where $\Delta N$ is the population inversion, n is the refractive index, g(v) is a normalized line shape function.[1]

The output power from a laser is

$$P_0 = \frac{V_m I_s}{l}\left(\frac{\gamma l}{L_i + T} - 1\right)T \qquad (S3)$$

where $V_m$ is mode volume, $T$ is intensity transmission, $l$ is round trip distance, $\gamma l$ is round trip gain, $L_i$ is round trip loss, $I_s$ is the saturation intensity. The population inversion $\Delta N$ will increase with increased pumping, the round-trip gain will also increase. Therefore, the output power can be increased by using higher pumping density.

Since MAPbI$_3$ has a much larger absorption coefficient than that of PbI$_2$, more pumping power will reach the inner layer of the nanoplatelet as the MAPbI$_3$ degrades to PbI$_2$ according to lambert-beer law of linear absorption

$$I = I_0 \exp(-\alpha z) \qquad (S4)$$

where $\alpha$ is the linear absorption coefficients.

Therefore, the more MAPbI$_3$ molecules in the inner layer of the nanoplatelet will contribute to the population inversion at the beginning of the degradation. However, the population inversion cannot be sustained anymore with more MAPbI$_3$ degrades.

2. Transient thermal response of a MAPbI$_3$ nanoplatelet

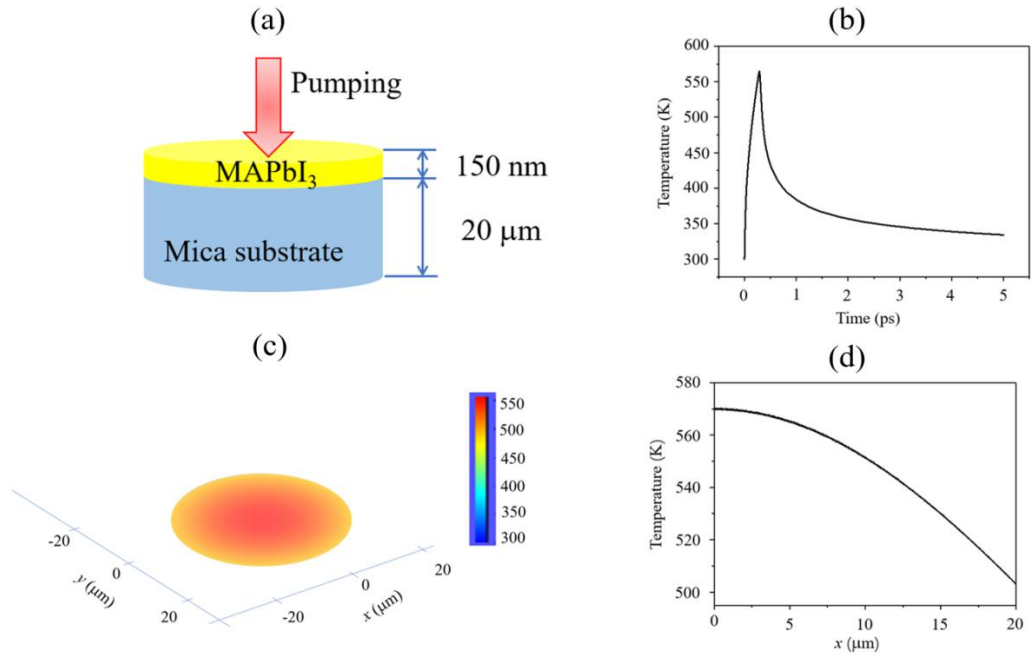

**Figure S6.** (a) Schematic diagram of a MAPbI$_3$ nanoplatelet on mica substrate being heated by a pump laser; (b) Transient thermal response of a MAPbI$_3$ nanoplatelet; (c) Temperature of a MAPbI$_3$ nanoplatelet at 290 fs after being pumped by a 290 fs laser pulse; (d) Radial temperature distribution a MAPbI$_3$ nanoplatelet at 290 fs after being pumped by a 290 fs laser pulse.

A three-dimensional heat transfer model is solved by finite difference method to determine the time-dependent temperature distribution in the perovskite nanoplatelets. The hexagonal MAPbI$_3$ nanoplatelet is simplified to be a round-shape nanoplatelet with a thickness of 150 nm and diameter of 40 μm as shown in Figure S5a. It is heated for 290 fs by a 343 nm pump laser with a peak power of 6666.7 W (Pump density of ~17 μJ/cm$^2$). The laser beam has a Gaussian intensity profile with a beam diameter of 107 μm. Since MAPbI$_3$ nanoplatelet has much higher absorption at 343 nm than that at 780 nm, the 343 nm pump laser is modeled as the only heat source. Thanks to the exceptional large absorption coefficient of MAPbI$_3$, most of pumping light energy is absorbed by the nanoplatelet and the optical penetration depth is ~15 nm. Therefore, the absorption at the mica substrate is neglected. The pump laser has Gaussian-shape beam profile with average power of 11.5 μW, beam radius of 107 μm, pulse duration time of 290 fs. The environment temperature is set as the same as 300 K.

**Table S1.** Parameters of the materials used for transient thermal response simulation.

|  | Absorption coefficient [cm$^{-1}$] | Thermal conductivity [W m$^{-1}$ K$^{-1}$] | density [kg/m$^3$] | Heat capacity [J K$^{-1}$ kg$^{-1}$] |
|---|---|---|---|---|
| MAPbI$_3$ | 6×10$^5$(2) | 0.5(3) | 3947(4) | 241.9(3) |
| Mica | – | 0.75(5) | 2900[6] | 880(5) |

Figure S5b shows transient thermal response of MAPbI$_3$ nanoplatelet. As can be seen, the temperature of the nanoplatelet increases to 570 K in 290 fs and gradually falls down to 5 ps. Therefore, MAPbI$_3$ suffers a significant temperature increase (270 K) in a short time which induces the nanoplatelet to degrade gradually from the dangling bonds.

The temperature distribution of the nanoplatelet at 290 fs is shown in Figure S5c. It can be seen that the center of nanoplate has higher temperatures as a result of the Gaussian beam profile. The temperature distribution along the radial direction is shown in Figure S5d. As can be seen, the temperature of the edge of the nanoplatelet is still higher than 500 K which is also high enough to induce the thermal degradation of the nanoplatelet.

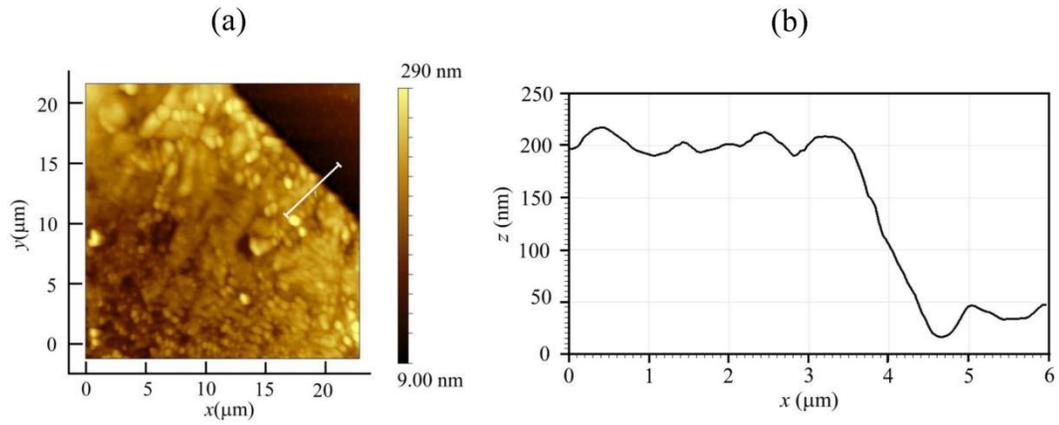

**Figure S7.** AFM image of the edge of PbI$_2$ passivated MAPbI$_3$ nanoplatelet (a) and corresponding thickness.

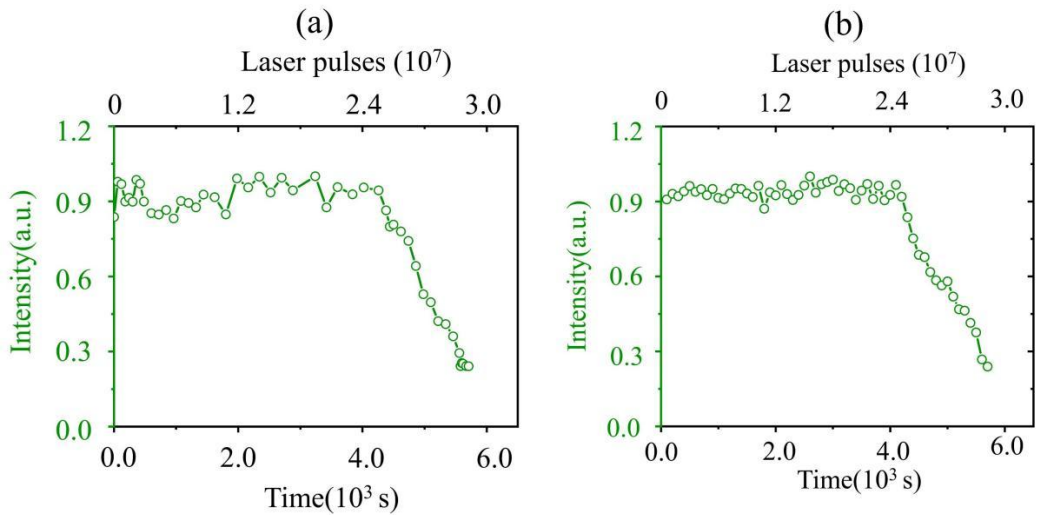

**Figure S8.** Lasing stability data of another two PbI$_2$ passivated MAPbI$_3$ lasers (a) and (b) under femtosecond laser pumping with a repetition rate of 6 kHz in ambient air condition.

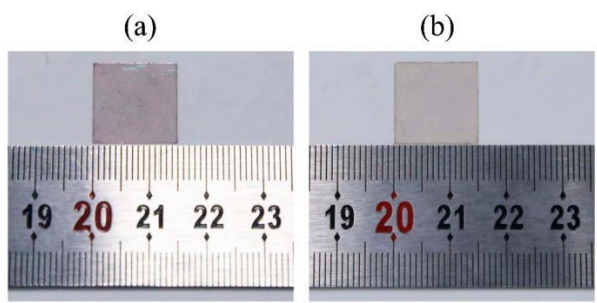

**Figure S9.** (a) Image of PbI$_2$ passivated MAPbI$_3$ nanoplatelets encapsulated with a DBP film on mica substrate and (b) Image of unpassivated MAPbI$_3$ nanoplatelets on mica substrate.

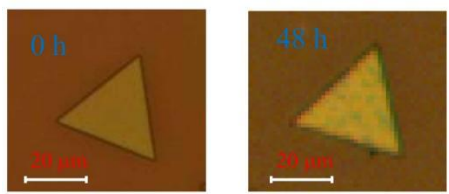

**Figure S10.** Microscopic images of a MAPbI$_3$ nanoplatelet after leaving in ambient air condition for 0 h (a) and 48 h (b), respectively.

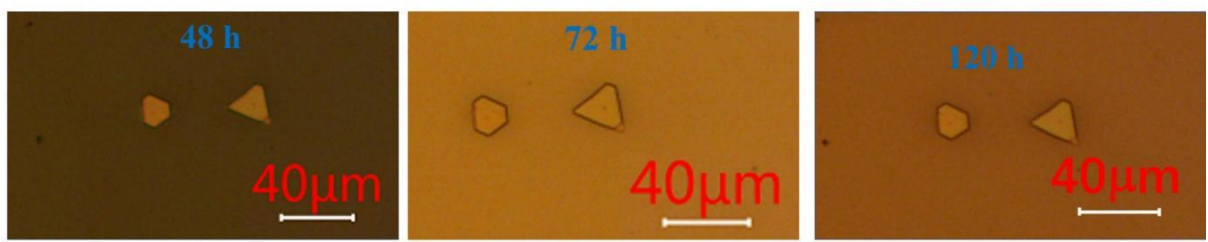

**Figure S11.** Microscopic images of PbI$_2$ passivated MAPbI$_3$ nanoplatelets encapsulated with a DBP film after leaving in ambient air condition for different times.

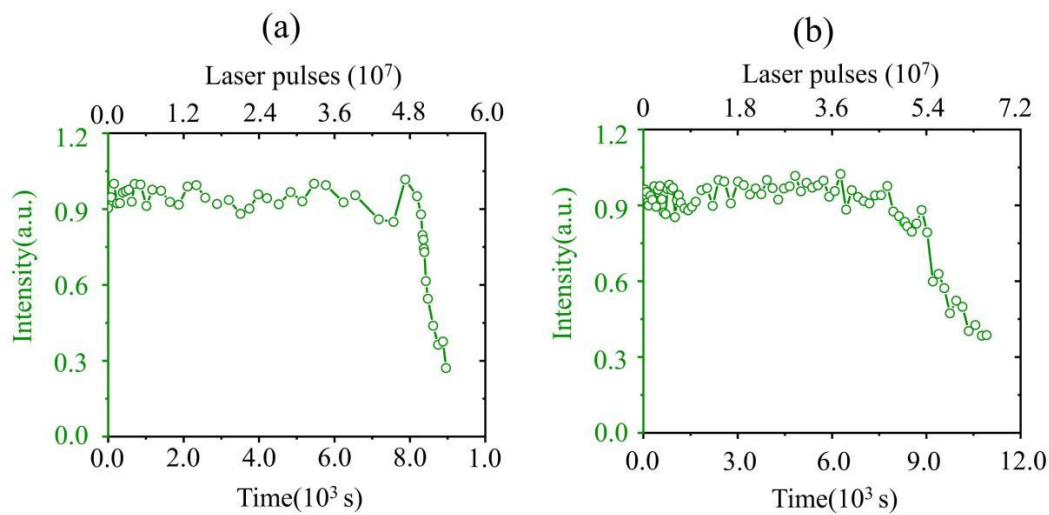

**Figure S12.** Lasing stability data of another two dual passivation processed MAPbI$_3$ lasers (a) and (b) under femtosecond laser pumping with a repetition rate of 6 kHz in ambient air condition.

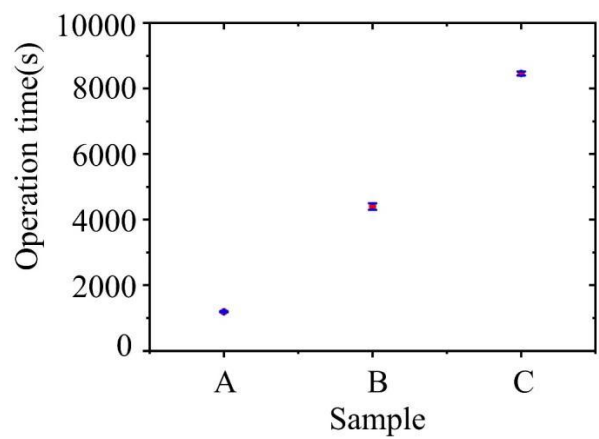

**Figure S13.** Average operation time of unpassivated (sample A), PbI$_2$ passivated (sample B) and dual passivation processed nanoplatelet lasers (sample C) under femtosecond laser pumping with a repetition rate of 6 kHz in ambient air condition.

Reference

1. Nagourney, WG. Quantum Electronics for Atomic Physics2010 6/11/2010.
2. Wang, Y.; Zhang, Y.; Zhang, P.Zhang, W.   High Intrinsic Carrier Mobility and Photon Absorption in the Perovskite Ch3nh3pbi3. *Phys Chem Chem Phys* **2015**,17,11516-20.
3. Jia, Y.; Kerner, RA.; Grede, AJ.; Rand, BP.Giebink, NC.   Factors That Limit Continuous-Wave Lasing in Hybrid Perovskite Semiconductors. *Advanced Optical Materials* **2020**,8,1901514.
4. Birowosuto, MD.; Cortecchia, D.; Drozdowski, W.; Brylew, K.; Lachmanski, W.; Bruno, A.Soci, C.   X-Ray Scintillation in Lead Halide Perovskite Crystals. *Scientific Reports* **2016**,6,37254.
5. Wang, S.; Ai, Q.; Zou, T-q.; Sun, C.Xie, M.   Analysis of Radiation Effect on Thermal Conductivity Measurement of Semi-Transparent Materials Based on Transient Plane Source Method. *Applied Thermal Engineering* **2020**,177,115457.
6. [https://www.tedpella.com/vacuum_html/Mica_Grade_V1_Properties.html](https://www.tedpella.com/vacuum_html/Mica_Grade_V1_Properties.html).